\documentclass[prb,twocolumn,showpacs,superscriptaddress,amsmath,amssymb]{revtex4}
\def\LamF{{\lambda_{\rm F}}}
\def\Kf{{k_{\rm F}}}
\def\Ef{{k_{\rm F}^{2}}}
\def\Hand{\hat{H}}

\newcommand{\eqbreak}{
\end{multicols}
\begin{widetext}
\noindent
\rule{.48\linewidth}{.1mm}\rule{.1mm}{.1cm}
}
\newcommand{\eqresume}{
\noindent
\rule{.52\linewidth}{.0mm}\rule[-.1cm]{.1mm}{.1cm}\rule{.48\linewidth}{.1mm}
\begin{multicols}{2}
\end{widetext}
}

\newcommand{\kf}{k_{\rm F}}
\newcommand{\ef}{k_{\rm F}^{2}}

\newcommand{\brho}{{\bm{\rho}}}
\newcommand{\BdG}{Bogoliubov--de~Gennes}
\newcommand{\myx}{s}
\newcommand{\mys}{\sigma}
\input{epsf}
\usepackage{graphicx}
\usepackage{dcolumn}
\usepackage{bm}
\begin{document}
\title{Density of states in d-wave superconductors
disordered by extended impurities}
\author{\.{I}nan\c{c}~Adagideli}
\email{adagidel@lorentz.leidenuniv.nl}
\affiliation{
Instituut-Lorentz, Universiteit Leiden,
Niels Bohrweg 2, Leiden, NL-2333 CA, The Netherlands}
\author{Daniel E.~Sheehy}
\email{sheehy@physics.ubc.ca}
\affiliation{
Department of Physics \&\ Astronomy,
Univ.~of British Columbia,
6224 Agricultural Rd., Vancouver, B.C.~V6T1Z1, Canada}
\author{Paul M.~Goldbart}
\email{goldbart@uiuc.edu}
\affiliation{
Department of Physics,
University of Illinois at Urbana-Champaign,
1110 West Green Street,
Urbana, Illinois 61801, U.S.A.}

\date{January 18, 2002}
\begin{abstract}
The low-energy quasiparticle states of a disordered d-wave superconductor
are investigated theoretically.  A class of such states, formed via tunneling
between the Andreev bound states that are localized around extended impurities
(and result from scattering between pair-potential lobes that differ
in sign) is identified.   Its (divergent) contribution to the total density
of states is determined by taking advantage of connections with certain
one-dimensional random tight-binding
models.  The states under discussion should be distinguished from those
associated with nodes in the pair potential.
\end{abstract}
%
%
%
%
%
\maketitle
\noindent
{\sl Introduction\/}:
In recent years, considerable attention has been focused on the low-energy
electron-hole quasiparticle spectral properties of the cuprate
superconductors in the presence of impurity scattering.  Much of the impetus
for this effort has its origin in the fact that
many of the cuprate superconductors are randomly chemically doped
insulators, and are therefore disordered. Moreover, as they are
pair-breakers for them, the role of impurities is especially
important for d-wave superconductors. Of particular interest is
the behavior of the single-particle density of states (DOS)
$\rho(E)$ as the energy $E$ tends to zero, i.e., its low-energy
behavior.

In recent work on the DOS of disordered d-wave superconductors,
P\'epin and Lee~\cite{REF:Pepin} invoked a $t$-matrix
approximation to infer that $\rho(E)\sim 1/E|\ln E^2|^2$ at low energies.  More
recently, Yashenkin et al.~\cite{REF:Yashenkin01} and
Altland~\cite{REF:Altland} have argued that the divergence found
in Ref.~\cite{REF:Pepin} is present only for the case of a
vanishing chemical potential (i.e.~for a half-filled band), and thus
does not apply to a doped cuprate.  (To be precise, unitarity of
the impurity scattering is also required.)\thinspace\  It is
further argued in Refs.~\cite{REF:Yashenkin01,REF:Altland} that,
instead of diverging, $\rho(E)$ should vanish at $E=0$ (unless
certain very specific fine-tuning requirements are met).
An important feature shared by
Refs.~\cite{REF:Pepin,REF:Yashenkin01,REF:Altland} is the
hypothesis that the disorder potential may be adequately modeled
by a random collection of {\it point-like\/} scatterers.
However,
for a {\it single\/} impurity in a d-wave superconductor, the
low-energy DOS is qualitatively different for
point-like~\cite{REF:Balatsky} and extended~\cite{REF:Inanc}
(i.e. impurities of a size much larger than the Fermi wavelength)
impurities: the states that reside at zero energy for extended
impurities reside at nonzero energies for point-like impurities.
(The underlying reason for this difference is that
for point-like impurities, the quasiparticle scattering is
essentially diffractive, whereas for the extended impurities it is essentially
semiclassical.) This observation raises the possibility that such differences will
continue to manifest themselves in the many-impurity setting.

The purpose of the present Paper is to identify a mechanism for
producing low-energy quasiparticle states. This mechanism is based on
impurity-scattering processes that connect states associated with
differing signs of the d-wave pair potential~\cite{REF:signchange}. 
In the case of a single
impurity~\cite{REF:Inanc}, this mechanism has already been shown
to produce low-energy states that are localized near the impurity.
These states can be associated with the classical trajectories scattering
from the impurity, and have been observed via scanning-tunneling
spectroscopy~\cite{REF:Yazdani,REF:JCDavis}.

Here, we build upon this single-impurity physics to identify a singular
(and potentially dominant) contribution to the low-energy DOS.  This
contribution, which conventional techniques fail to capture, arises from
tunnelling along the classical trajectories that connect the individual
impurities and, hence, connect the low-energy states localized near these
impurities.  The underlying physics was formulated some time ago in the
context of tunnelling corrections to ground-state energies in in models of supersymmetric
quantum mechanics~\cite{REF:Witten81,REF:SvH82,REF:Junker}.

\begin{figure}[hbt]
 \epsfxsize=2.9in
\centerline{\epsfbox{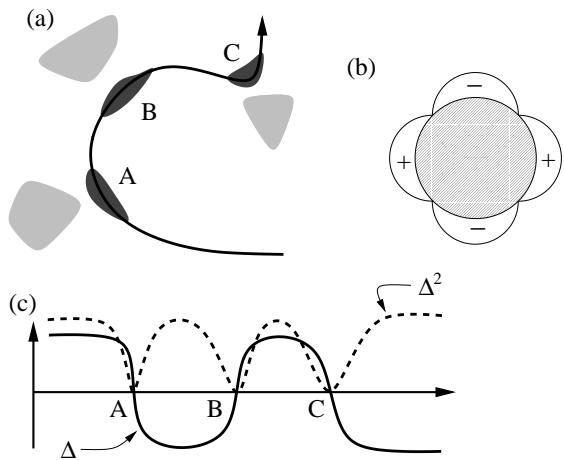}}
\vskip0.50cm \caption{ (a) Sketch
of a classical trajectory that encounters several impurities and,
hence, several sign-changes in $\Delta$.  The shaded regions on
the trajectory denote the sections of the trajectory where
approximate zero-energy states reside. (b)~Momentum-space pair
potential of a d-wave superconductor. (c)~Schematic depiction of
$\Delta$ and $\Delta^2$ along the trajectory.
 }
\label{FIG:scatter}
\end{figure}
The picture we have in mind(see Fig.~\ref{FIG:scatter}) of the processes
that lead to low-energy states involves classical trajectories that each
visit many extended impurities.
As a quasiparticle progresses along such a trajectory,
its momentum is repeatedly altered via scattering from the extended
impurities, so that the effective pair-potential
 (Fig.~\ref{FIG:scatter}b)
undergoes
sign changes.  Localized near each
such sign-change would be a zero-energy quasiparticle state; quasiparticle
tunnelling (through the pair-potential) connects these states, raising
their energies from zero, and thus forming a low-energy band that exhibits a
Dyson-like singularity~\cite{REF:Dyson} at zero energy:
$\rho(E)\sim 1/E\vert\ln E^2\vert^{3}$.
This picture loses its precision for sequences of impurities
between which the pair potential is predominantly small (i.e.~for nodal
directions) and, consequently, the states are not well localized near the impurities.
However, the contribution on which we are focusing (i.e.~the non-nodal
contribution) is expected to be substantial, and quite likely dominant,
in the low-energy limit.


Based on previous work on low-energy quasiparticle
states~\cite{REF:Inanc,REF:Balatsky}, we identify three classes
of processes that can effect the low-energy DOS of a disordered
d-wave superconductor: (i)~purely semiclassical scattering
between states with differing signs of the pair-potential
(i.e.~scattering due to random extended scatterers in d-wave
superconductors), which we shall focus on in the present Paper;
(ii)~purely diffractive scattering between states near the nodal
directions of the pair-potential (i.e.~the scattering of nodal
quasiparticles by point-like impurities), which was considered
in, e.g., Refs.~\cite{REF:Altland,REF:Yashenkin01,REF:Chamon01};
and (iii)~the mixing of the low-energy states that arise from
processes~(i) and (ii). 
Throughout the rest of this Paper
we shall ignore processes in class~(iii). In consequence, the
low-energy DOS of a disordered d-wave superconductor can be
expressed as a sum of contributions arising from class~(i)
processes (viz.~$\rho_{\rm ext}$) and class~(ii) processes
(viz.~$\rho_{\rm nodal}$), the latter, as discussed in
Refs.~\cite{REF:Yashenkin01,REF:Altland}, being non-divergent
and therefore subdominant.  Thus, we shall focus on $\rho_{\rm
ext}$ [and denote this by $\rho(E)$].

\noindent
{\sl Quasiparticle motion in a d-wave superconductor
with many extended impurities\/}:
Our focus will be on the DOS per unit area, i.e.,
\begin{equation}
\label{eq:rho}
\rho(E)\equiv{\frac{1}{A}}\sum_m \delta(E-E_m),
\end{equation}
where $A$ is the area of the sample and the energy
eigenvalues $\{E_m\}$ follow from the \BdG\ eigenproblem, viz.,
\begin{equation}
\label{eq:bdg}
\begin{pmatrix}
\hat{h} & \hat{\Delta} \cr
         \hat{\Delta} & -\hat{h}
\end{pmatrix}
\begin{pmatrix}
u_n \cr v_n
\end{pmatrix}
=E_n
\begin{pmatrix}
u_n\cr v_n
\end{pmatrix}.
\end{equation}
Here,
$\hat{h}\equiv-\nabla^{2}-\Ef+V({\bf r})$,
in which $\Ef$ is the chemical potential
[i.e.~$\Kf$ ($\equiv 2\pi/\LamF$) is the Fermi wave vector],
$V$ is the single-particle impurity potential,
and we have adopted units in which $\hbar^2/2m=1$,
$m$ being the common (effective) mass of the electrons and holes.
The operator $\hat{\Delta}$ is the pair-potential (integral) operator;
how it acts is specified by the nonlocal pair-potential kernel
$\Delta({\bf r},{\bf r}')$ via
$[\hat{\Delta}v_n]({\bf r})=
\int d^2 r'\Delta({\bf r},{\bf r}')\,v_n({\bf r}')$.

To define this model fully, we need an assumption about the form of
$\Delta({\bf r},{\bf r}')$.  It is convenient to exchange the coordinates
${\bf r}$ and ${\bf r}'$ for the relative and center-of-mass coordinates
$\brho$ and ${\bf R}$:
\begin{equation}
\bar{\Delta}(\brho,{\bf R})
\equiv
\Delta({\bf r},{\bf r}'),
\quad
\brho \equiv{\bf r}-{\bf r}',
\quad
2{\bf R}\equiv{\bf r}+{\bf r}'\,.
\end{equation}
Then, via Fourier transformation with respect to $\brho$, viz.,
\begin{equation}
\bar{\Delta}({\bf k},{\bf R})
\equiv\int d^2\!\rho\,
{\rm e}^{-i{\bf k}\cdot \brho}\,
\bar{\Delta}(\brho,{\bf R}),
\label{eq:deltaft}
\end{equation}
we obtain the pair-potential $\bar{\Delta}({\bf k},{\bf R})$
at center-of-mass position ${\bf R}$
and relative momentum ${\bf k}$.
As our aim is to describe the cuprate superconductors, we take
$\bar{\Delta}({\bf k},{\bf R})$ to have d-wave symmetry:
$\bar{\Delta}({\bf k},{\bf R}) \propto (k_x^2-k_y^2)$, where
$k_x$ and $k_y$ denote the cartesian components of ${\bf k}$.
However, we allow for the possibility of position-dependent amplitude
variations of the d-wave pair-potential due, say, to pair-breaking effects
near the extended scatterers.

\noindent
{\sl Semiclassical approach to the
Bogoliubov--de~Gennes eigenproblem\/}:
We now invoke a semiclassical approximation under which
$\rho(E)$ is expressed in terms of the solution of
a family of one-dimensional eigenproblems, each associated with a classical
scattering trajectory in the presence of the single-particle impurity
potential $V({\bf r})$.  We restrict ourselves to a brief discussion of
this approach; for details see Refs.~\cite{REF:Inanc,REF:Sheehy}.
The approximation amounts to our:
  (i)~regarding the kinetic and potential energies as being comparable
      and being the largest energies in the problem;
 (ii)~turning off the pair potential;
(iii)~treating semiclassically the quasiparticle motion in the
      presence of the kinetic and potential energies; and
 (iv)~reinstating the pair potential.
Via this approach, we reduce the two-dimensional \BdG\ eigenproblem to
a family of one-dimensional Andreev eigenproblems residing on trajectories,
each trajectory being a particular classical scattering
trajectory in the presence of the many-impurity potential.
This scheme applies under the following conditions:
 (i)~the amplitude of $\bar{\Delta}$ and $V$ should vary slowly,
relative to the Fermi wavelength $\lambda_{\rm F}$; and
(ii)~the Fermi energy $k_{\rm F}^2$ should be large compared with the
energy scale of interest, viz.~$E$, as well as with the typical pair-potential
scale.

Let us now turn to the family of one-dimensional eigenproblems arising
from this semiclassical scheme.  Following Ref.~\cite{REF:Inanc}, these
trajectory-dependent eigenproblems take the form
\begin{subequations}
\begin{eqnarray}
\label{EQ:2ndOBdG}
&&\Hand
\begin{pmatrix}
\bar{u}_n \cr \bar{v}_n
\end{pmatrix}
=
E_n
\begin{pmatrix}
\bar{u}_n \cr \bar{v}_n
\end{pmatrix},
\\
&&\Hand
\equiv
\begin{pmatrix}
-2i\Kf\,\partial_{\mys} &
\Delta_0(\mys)\cr
\Delta_0(\mys)&
2i\Kf\,\partial_{\mys}
\end{pmatrix},
\label{EQ:2ndOBdG2}
\\
&&\Delta_0(\mys)
\simeq
\bar{\Delta}
\big(\kf\partial_{\mys}{\bf x}_{\rm c}(\mys),{\bf x}_{\rm c}(\mys)\big),
\label{eq:deltaeff}
\end{eqnarray}
\end{subequations}
i.e., an Andreev eigenproblem~\cite{REF:Andreev}.
Here, the parameter $\mys$ measures the position along a particular
classical trajectory ${\bf x}_{\rm c}(\mys)$, the latter obeying
Newton's equation in the many-impurity potential, viz.,
\begin{equation}
\ef\,\partial_{\mys}^{2}{\bf x}_{\rm c}(\mys)=
-{\bm{\nabla}}V({\bf x}_{\rm c}(\mys)).
\end{equation}
The DOS is then obtained by assembling the eigenvalue
spectra $E_n({\bf n},b)$ of all the classical trajectories, the latter
being labelled in terms of an asymptotic momentum direction ${\bf n}$
and impact parameter $b$:
\begin{subequations}
\begin{eqnarray}
\label{eq:rhofin}
\rho(E)&\simeq&
{\frac{\kf}{A}}\int\frac{d{\bf n}}{2\pi}\,
\int db\,\rho({\bf n},b,E),\\
\label{eq:rhotraj}
\rho({\bf n},b,E)
&\equiv&
\sum_m\delta
\big(E-E_m({\bf n},b)\big).
\end{eqnarray}
\end{subequations}
Thus, in order to obtain $\rho(E)$ one needs to find
each classical trajectory, obtain the associated effective pair-potential
[given by Eq.~(\ref{eq:deltaeff})], solve the resulting one-dimensional eigenvalue
equation and, finally, integrate over all the classical trajectories using
Eq.~(\ref{eq:rhofin}).
We note that
if one interprets the weight of a particular classical trajectory as the
probability of finding a pair-potential configuration corresponding to that
particular trajectory then we see that the calculation of $\rho(E)$ amounts to
computing the average density of states of a random pair-potential model.
Models
of this sort have been considered, e.g., in
Refs.~\cite{REF:Waxman,REF:Comtet,REF:LB-MPAF97,REF:Bartosch}.

\noindent
{\sl Eigenvalue problem for a single trajectory\/}:
We now examine the contribution $\rho({\bf n},b,E)$ to the DOS for
the case of a generic trajectory $({\bf n},b)$.  For convenience,
we introduce the rescaled trajectory parameter
$\myx\equiv\mys/2\kf$;  the Hamiltonian then becomes
\begin{equation}
 \Hand=
\begin{pmatrix}
-i\,\partial_{\myx}&
\Delta(\myx)\cr
\Delta(\myx)&
i\,\partial_{\myx}
\end{pmatrix},
\end{equation}
where $\Delta(\myx)\equiv\Delta_0(2\kf\myx)$.

Our method for calculating the spectrum of $\Hand$ in the many-impurity
case is based on that for a single impurity~\cite{REF:Inanc}.
In the latter case, low-energy states arose from asymptotically
sign-changing trajectories [i.e.~those trajectories for which
$\lim_{\myx\to\pm\infty}\Delta(\myx)$ differ in sign].
Finding the spectrum amounted to identifying such sign-changing trajectories.
What about the
case of {\it many\/} extended impurities?  In this case, for a typical
trajectory through the impurity potential
$\Delta(\myx)$ undergoes repeated sign changes.  On a particular
trajectory let us label the the positions of these zeroes of $\Delta(\myx)$
by $\{\myx_n\}$.
Recall that we are concerned with the collection of impurity
states that would lie at zero energy if the impurities were
isolated.  Owing to tunnelling between them, these formerly
degenerate states yield a continuum of states, extending upwards
in energy from zero.  Our task is to shed some light on this
band-formation. We proceed to set up a tight-binding model along
the trajectory, in which we retain only the zero-energy impurity
states $\{\vert{n}\rangle\}$ (i.e.~the local ground states at each of the
$\{\myx_{n}\}$) and allow only
nearest-neighbor tunneling between them~\cite{REF:corrections,REF:instanton}.
 To complete the model,
we need the matrix elements of $\Hand$ connecting these
states, i.e.
\begin{equation}
t_{n}\equiv \langle{n}\vert \Hand \vert{n+1}\rangle.
\end{equation}
Using the analytic expression for the zero energy wavefunctions~\cite{REF:Junker},
$
\langle{s}\vert{n}\rangle
\propto
\exp -\int^{s}_{s_n}ds'\,\vert\Delta(s')\vert
$,
a straightforward calculation produces
\begin{equation}
\label{eq:hop-me} t_n \!\approx\! \frac{1}{\sqrt{\pi}}
|\Delta'(\myx_n)\Delta'(\myx_{n+ 1})|^{1/4} \exp\Big\{
-\int\nolimits_{\myx_n}^{\myx_{n+1}}{\!\!\!}d\myx' |\Delta(\myx')|\Big\},
\end{equation}
where $\Delta'(\myx)\equiv\partial_\myx\,\Delta(\myx)$.
We are now in a
position to write down a low-energy effective approximation to
$\Hand$, viz.,
\begin{equation}
\label{eq:htunn}
\Hand\approx\sum_n t_n
\big(| n  \rangle \langle n+1 | +
     | n+1\rangle \langle n   |\big),
\end{equation}
i.e., for each classical trajectory, one arrives at a (topologically)
one-dimensional hopping model that captures the physics of tunneling
processes between the (formerly zero-energy) states localized near each
zero of the pair-potential.

\noindent
{\sl Density of states\/}:
In order to obtain the low-energy DOS, we must obtain the DOS of the effective
Hamiltonian~(\ref{eq:htunn}) for each trajectory, and then collect them together.
We assume that the collection of trajectories forms an ensemble that is characterized
by the condition that momentum directions before and after a collision are uncorrelated.
Then, summing over such an ensemble of trajectories is equivalent (up to a constant of
proportionality) to averaging the DOS of the Hamiltonian~(\ref{eq:htunn}) over
uncorrelated values of $t_n$.
To obtain the low-energy DOS of this
effective model we appeal to results obtained by Eggarter
and Riedinger~\cite{REF:ER78}, who, building on the work of
Dyson~\cite{REF:Dyson} and Theodorou and Cohen~\cite{REF:TC76},
studied random-hopping models of precisely this form.
Specifically, in Ref.~\cite{REF:ER78} it was found that, under the condition that
the $\{ t_n \}$ are uncorrelated~\cite{REF:uncorrelated} and identically
distributed, the DOS as $E\rightarrow 0$ is given by
\begin{equation}
\label{eq:doscontr}
\rho(E)
\approx N_{\rm s}\,Z {\frac{2\sigma^2}{E\,\vert\ln (E/\bar{t}\,)^2\vert^{3}}},
\end{equation}
where $N_{\rm s}$ denotes the average number of sites along the trajectory,
$Z$ is the constant of proportionality arising from the Jacobian of the 
transformation from summing over trajectories to averaging over $\{t_n\}$,
$\bar{t}$ is the scale characterizing $\{t_n\}$,
and the amplitude $\sigma^2$ is given by the variance of the
logarithm of $t$, i.e.,
\begin{equation}
\sigma^2\equiv
\big\langle (\ln t^2/\Delta_0^2 )^2 \big\rangle -
\big\langle \ln t^2/\Delta_0^2 \big\rangle^2\,,
\end{equation}
where $\langle\cdots\rangle$ denotes a disorder average.
The scale $\bar{t}$ and $\sigma^2$ could be estimated with the help 
of Eq.~\ref{eq:hop-me} as
\begin{subequations}
\begin{eqnarray}
&&\bar{t}^2\propto k_{\rm F} \Delta_0 n_{\rm c}^{1/2}
\exp(-\Delta_0/k_{\rm F}n_{\rm c}^{1/2}),
\\
&&\sigma^2\propto\Delta_0/k_{\rm F}n_{\rm c}^{1/2},
\end{eqnarray}
\end{subequations}
where $n_{\rm c\/}$ is the number of impurities per unit area.
What remains is to determine the coefficient $Z$; we now make
an estimate of this quantity.

Our estimate for $Z$ follows from considering the integration 
over $b$ for a single impurity of size $a$.  
As the impurity potential is expected to decay rapidly away from the 
impurity, the only trajectories that interact appreciably with it are
those that directly intersect it, i.e., $\kf \int db \to \kf \,a$.
Thus, in the absence of the tunneling corrections between the 
zero-energy
states, $\rho(E)$ is approximately given by
$\rho_{0}(E)\approx\kf \,a \,n_{\rm c\/}\,\delta(E)$~\cite{REF:Inanc}.
Although the inclusion of tunneling corrections
changes the energy dependence of $\rho(E)$, here we assume that
the energy-integrated density of states $\int_0^{\epsilon}\rho(E)\,dE$ is
approximately conserved for some appropriately chosen cutoff $\epsilon$.
These considerations
lead to the following approximate form for $\rho(E)$ valid for $E\ll\bar{t}$:
\begin{equation}
\label{eq:final}
\rho(E) \propto {\frac{\Delta_0\, a \,n_{\rm c}^{1/2}}{E\,\vert\ln (E/\bar{t}\,)^2\vert^{3}}}.
\end{equation}

We remark that the divergence of the low-energy DOS is ultimately cut-off due to
physical processes not included in the present description. These include the dephasing
scale $\hbar/\tau_\phi$ and the diffractive scattering scale $\hbar/\tau_{\rm d.s.}$; the
cut-off will occur at the largest of these scales. 
Thus, it is possible that at extremely low energies $\rho(E)$ eventually does vanish
asymptotically, in agreement with the results of Refs.~\cite{REF:Yashenkin01,REF:Altland}. 
However, the present results would still apply at intermediate energies.   
Finally,
we stress that the calculation presented here does not place any
special emphasis
on the nodes of the d-wave order parameter; indeed, the singular
contribution to the
density of states of a d-wave superconductor reported here arises from states in
generic (rather than nodal) regions on
the Fermi surface.

\noindent
{\sl Acknowledgments\/}:
PMG thanks Alexei Tsvelik for
informative discussions and the University of Colorado at Boulder for its kind hospitality.
This work was supported by
the U.S.~Department of Energy
under Award No. DEFG02-ER9645439,
through the Frederick Seitz
Materials Research Laboratory,
by the Dutch Science Foundation NWO/FOM (IA),
 and by NSERC (DES).


\begin{thebibliography}{}
\bibitem{REF:signchange}
Of course, the mechanism for creating low-energy states invoked 
in the present Paper does not specifically require d-wave 
superconductivity.  All that is required is that the pair 
potential be real (or can be made so by a suitable gauge 
transformation) and that it possess regions of momentum 
orientation over which it has opposing signs.
\bibitem{REF:corrections}
Higher-order tunneling matrix elements are small, relative to the
leading-order contributions, provided that
$\int_{\myx_n}^{\myx_{n+1}} |\Delta(\myx)| d\myx \gg 1$~\cite{REF:SvH82},
i.e., that the coherence length is small compared with the typical
impurity spacing.
In addition, we note that, as $\Hand$ may be made purely off-diagonal by a suitable unitary
transformation, the tunneling amplitude between next-nearest neighbors vanishes
identically.
\bibitem{REF:instanton}
The related task of computing tunnelling corrections to
the ground state of a SUSY Hamiltonian having  multiple degenerate minima
using instanton methods has been considered in Ref.~\cite{REF:SvH82}.
\bibitem{REF:uncorrelated}
We suspect that the result for the DOS presented here will continue
to hold not solely in the setting of uncorrelated hopping matrix elements
$\{ t_n \}$ but also for correlations of finite range.
\end{thebibliography}
\end{document}